\documentclass[a4paper,aps,prl,twocolumn,superscriptaddress,showpacs]{revtex4}
\usepackage{graphicx}
\usepackage{amsmath}
\usepackage{amsfonts}
\usepackage{color}
\usepackage{bm}
\usepackage{pdfcomment}
\usepackage{pgfplots}
\usepackage{subfigure}
\usepackage{soul}

\newcommand{\br}{\boldsymbol{r}}

\newcommand{\beq}{\begin{equation}}
\newcommand{\eeq}{\end{equation}}

\begin{document}


\title{Entangling two defects via a surrounding crystal}

\author{T. Fogarty}
\affiliation{Physics Department, University College Cork, Cork, Ireland}
\affiliation{Quantum Systems Unit, Okinawa Institute of Science and Technology, Okinawa 904-0495, Japan}
\author{E. Kajari}
\affiliation{Theoretische Physik, Universit\"at des Saarlandes, D-66123 
Saarbr\"ucken, Germany}
\author{B. G. Taketani}
\affiliation{Theoretische Physik, Universit\"at des Saarlandes, D-66123 
Saarbr\"ucken, Germany}
\author{A. Wolf}
\affiliation{Theoretische Physik, Universit\"at des Saarlandes, D-66123 
Saarbr\"ucken, Germany}
\author{Th. Busch}
\affiliation{Physics Department, University College Cork, Cork, Ireland}
\affiliation{Quantum Systems Unit, Okinawa Institute of Science and Technology, Okinawa 904-0495, Japan}
\author{Giovanna Morigi}
\affiliation{Theoretische Physik, Universit\"at des Saarlandes, D-66123 
Saarbr\"ucken, Germany}
\affiliation{Departament de Fis\'ica, Universitat Aut\`onoma de Barcelona, 
E-08193 Bellaterra, Spain}

\date{\today}

\begin{abstract}
We theoretically show how two impurity defects in a crystalline structure can be entangled through coupling with the crystal. We demonstrate this with a harmonic chain of trapped ions in which two ions of a different species are embedded.  Entanglement is found for sufficiently cold chains and for a certain class of initial, separable states of the defects. It results from the interplay between localized modes which involve the defects and the interposed ions, it is independent of the chain size, and decays slowly with the distance between the impurities. These dynamics can be observed in systems exhibiting spatial order, viable realizations are optical lattices, optomechanical systems, or cavity arrays in circuit QED. 
\end{abstract}


\maketitle
\par

Entanglement is a quantum mechanical property with no classical counterpart. One of its peculiar features are the correlations between the measurement outcomes for two entangled objects, even when they are at large distances \cite{EPR}, which make entanglement a precious resource for quantum communication protocols and quantum metrology \cite{RMP_Horodecki}. On the other hand, its quantum mechanical nature makes it fragile against external perturbations \cite{RMP_Zurek}. 

The issue of identifying open-system dynamics in which entanglement is robust recently motivated a series of studies \cite{Mintert,Eberly:2004}, which showed that a bath can mediate entanglement between two objects interacting with it \cite{Lidar:1998,Braun:2002,Huelga:2002,Benatti:2003,Paz:2008,Galve:2010}. These findings lead to the question of how bath-mediated entanglement scales with the distance between the objects. When the bath is composed by oscillators forming a linear chain, entanglement between any pair of oscillators decays quickly on a length scale of the order of the interparticle distance \cite{Audenaert:2002,Anders:2008}. A further study, based on a phenomenological model, suggested that the cutoff wavelength, which in a linear chain coincides with the interparticle distance, is the characteristic length over which bath-mediated entanglement vanishes \cite{Zell:2009}. 

These predictions are however restricted to specific settings. In Ref. \cite{Ludwig:2010}, for instance, it was argued that Markovian models can not capture all dynamical processes which can lead to entanglement between distant objects. Moreover, in Ref. \cite{Wolf:2011} a microscopic model was proposed, that allowed two defect oscillators embedded in the chain to become entangled by the chain itself instead of thermalizing at the chain's temperature \cite{Rubin:1963}. This is possible since the impurities break the discrete translational symmetry of the lattice, giving rise to a set of normal modes involving the defects and the ions between them. Such modes are essential for the generation of entanglement at steady state. We remark that a linear chain of oscillators in a thermal state and coupled to a single impurity is a realization of Rubin's model, it leads to thermalization of the impurity and is an example of Brownian motion in a solid-state environment \cite{Rubin:1963}. The addition of a second impurity introduces a new symmetry modifying this behaviour. In the context of recent studies on non-Markovianity \cite{Wolf:2008,Ludwig:2010,Liu:2011,Vasile:2011,Huelga:2012} one can say that the second defect modifies the reservoir making it non-Markovian \cite{Wolf:2011}.

\begin{figure}[t]
\begin{center}
\includegraphics[angle=0,width=\columnwidth]{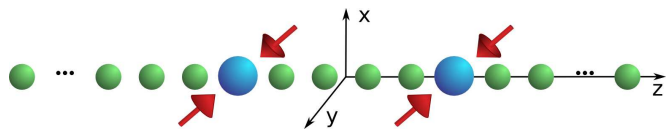}
\caption{(Color online) Two heavy ions are embedded in a linear chain of lighter ions. A standing-wave laser illuminates the defect ions and couples their transverse and axial displacements via the mechanical effects of light. For a certain class of initial separable states, the defects transverse  motion becomes entangled by the coupling with the axial phonons of the chain. \label{fig:model}}
\end{center}
\end{figure}

The findings of Ref. \cite{Wolf:2011} imply that two impurity defects can be entangled by the surrounding bulk, when the latter exhibits some sort of spatial order. In addition, entanglement could be found between two defects at any distance, independent of the size of the bulk. The latter statement thus goes beyond existing studies, which analyze entanglement generation between two systems at the edges of a chain \cite{Plenio:2004,Venuti:2006}. It further adresses the question, whether it is possible to generate entanglement between two macroscopically-distant objects through interaction with a large common bulk. Understanding this issue is relevant at the fundamental level, for example in the context of recent studies of thermalization in quantum systems \cite{Rigol:2008,Riera:2012}. Moreover, it is important for studies on biological systems \cite{Huelga:2012}, and for implementing protocols based on dissipative dynamics in quantum networks  \cite{Polzik:2011,Vollbrecht:2011,Gualdi:2011}, quantum computers \cite{Diehl:2008,Blatt:2011} and metrology \cite{Goldstein:2011,Chin:2012}.

In this Letter we analyse the dependence of bath-mediated entanglement on the distance by considering the specific physical system of two ions in a linear Paul trap, which are embedded in a linear chain of ions of different species, as shown in Fig.~\ref{fig:model}. Let $N$ and $Q$ be the number of ions and their charge. The ions are aligned along the $z$ axis inside a linear Paul trap with position (canonically-conjugated momentum)  ${\bf r_j}=(x_j,y_j,z_j)$ (${\bf p_j}=(p_{jx},p_{jy},p_{jz})$), where $j=1,\ldots,N$ labels the ions. The impurity defects are two ions of different species far away from the chain edges, whose mass $M$ is larger than the mass $m$ of the other ions composing the chain. The trap potential for the ion $j$ reads $V_{\rm trap}({\bf r_j})=(U_\parallel z_j^2+U_{\perp,j}(x_j^2+y_j^2))/2$, where $U_\parallel$ is determined by the static quadrupole potential and $U_{\perp,j}=(U_0/m_j-U_\parallel)/2$  by the radio-frequency potential creating the transverse confinement \cite{Kielpinski:2000}. We note the dependence of the transverse potential strength on the mass of the ion: This implies that the defect's transverse motion is a localized oscillation in the chain for sufficiently large ratios $M/m$. The ions are at a sufficiently low temperature $T$ such that they perform harmonic vibrations about their respective equilibrium positions $\br_j^{(0)}=(0,0,z_j^{(0)})$, at which the trap force and the Coulomb repulsion mutually balance. In this limit the dynamics is governed by the quadratic Hamiltonian $H_0=H_{\parallel}+H_{\perp}^{(x)}+H_{\perp}^{(y)}$, with
\begin{align}
&H_\parallel=\sum_{j=1}^N\left(\frac{p_{j,z}^2}{2m_j}+\frac{1}{2}U_\parallel q_j^2+\frac{1}{4}\sum_{\ell\neq j} \mathcal K_{j,\ell}(q_j-q_\ell)^2\right),\\
&H_\perp^{(x)}=\sum_{j=1}^N\left(\frac{p_{j,x}^2}{2m_j}+\frac{1}{2}U_{\perp,j}x_j^2-\frac{1}{8}\sum_{\ell\neq j} \mathcal K_{j,\ell}(x_j-x_\ell)^2\right),
\label{eq:hamiltonian}
\end{align}
and $H_{\perp}^{(y)}=H_{\perp}^{(x\to y)}$. Here, $q_j=z_j-z_j^{(0)}$, while $\mathcal K_{j,\ell}=2Q^2/|z_j^{(0)}-z_\ell^{(0)}|^3$ is the coupling due to the Coulomb repulsion \cite{Morigi:2004}. We label the defect ions by $j_1,j_2$, with $1\ll j_1<j_2\ll N$ and mutual distance $d\propto j_2-j_1\ll N$.

\begin{figure*}[t]
\begin{center}
\subfigure[]{\includegraphics[height=0.3\textwidth]{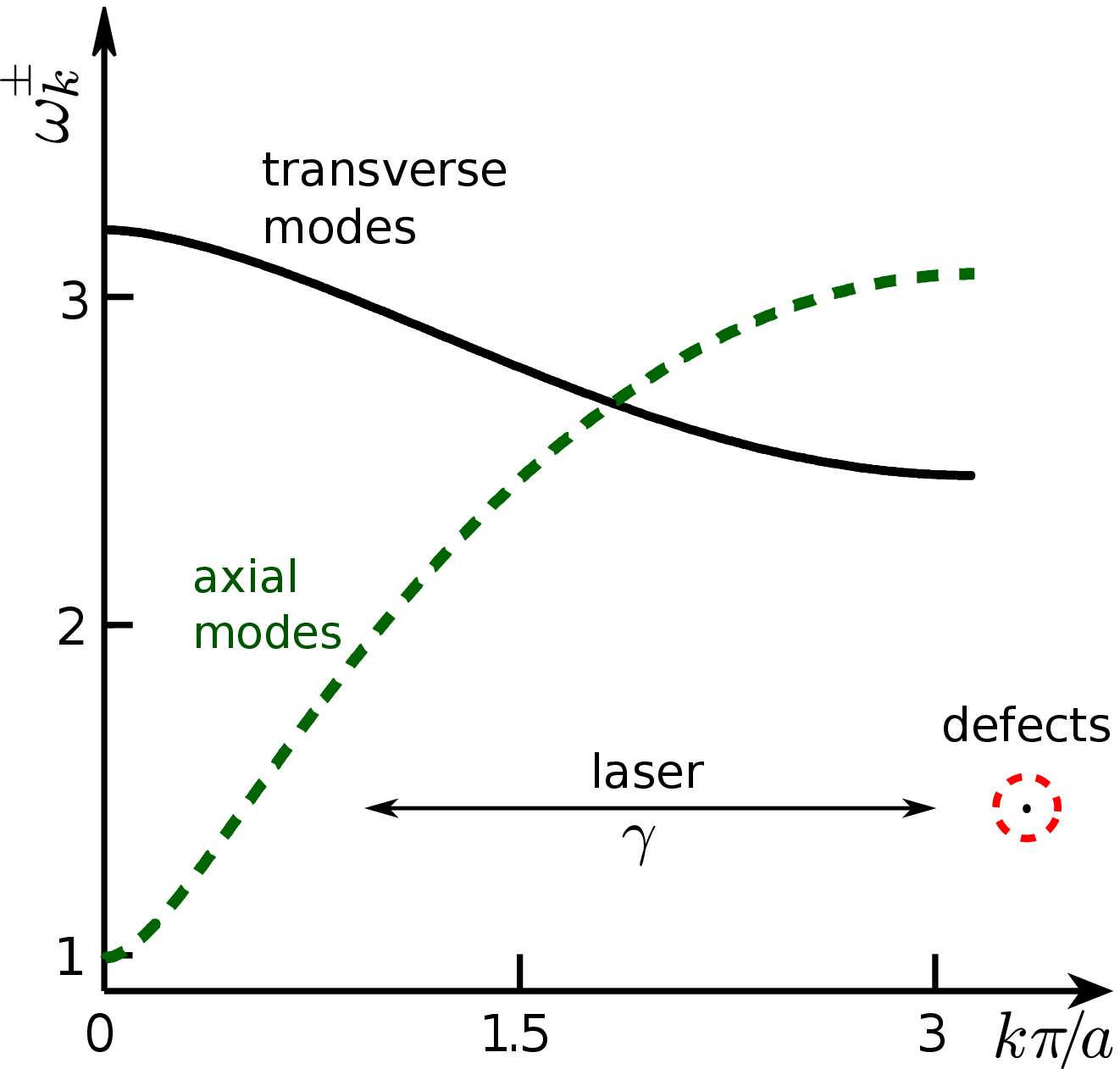}\label{fig:spectrum}}
\subfigure[]{\includegraphics[height=0.3\textwidth]{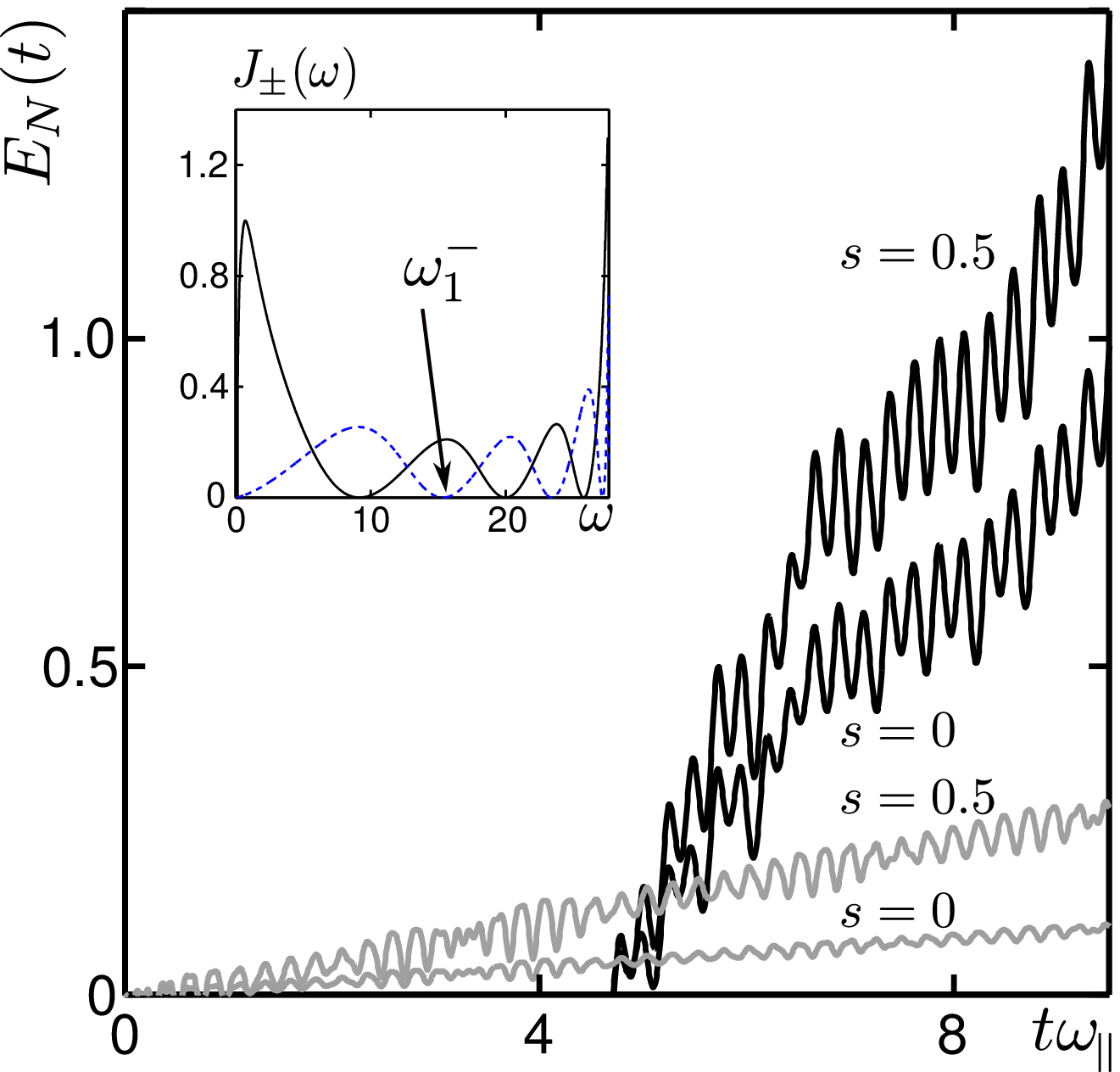}\label{fig:EN_dif_s}}
\hspace{0.3cm}
\subfigure[]{\includegraphics[height=0.3\textwidth]{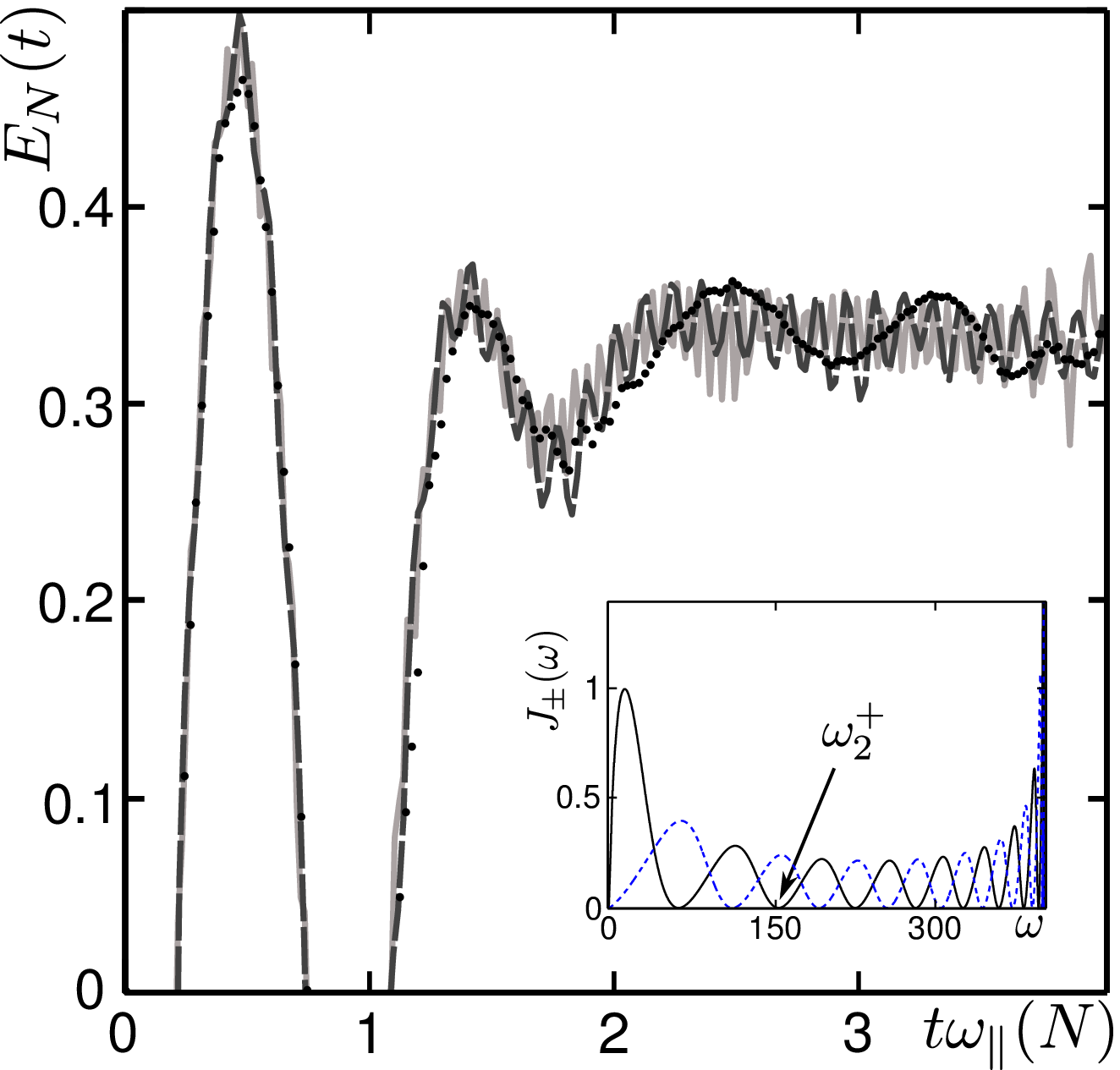}\label{fig:thermo_limit}}
\caption{(Color online) (a) Spectrum of excitations when  a chain of Ca$^+$ ions contains two In$^+$ ions with mutual distance $d=7a$. The mass ratio is $M/m\approx2.87$ \cite{Hayasaka}. The eigenfrequencies $\omega_k$ of $H_\perp$ (solid) and $H_\parallel$ (dashed line) are displayed as a function of the quasimomentum $k$ (in units of $\pi/a$). The isolated transverse frequencies, indicated by the circle, almost coincide with the frequency of the defects transverse oscillator. The parameters are $U_\perp(m)/U_\parallel=2000$ and $\omega_\parallel=2\pi\times 26.1$kHz, while the frequencies are in units of $\omega_\parallel=\sqrt{U_\parallel/m}$. (b) Logarithmic negativity $E_N$ as a function of time $t$ (in units of $\omega_\parallel^{-1}$) for $N=50$, initial squeezing $s=0, 0.5$,  while the rest of the chain is initially prepared at $T=8\hbar\omega_\parallel/k_B\approx 10\mu$K. The black (grey) curves represent the case with laser coupling switched on (off), with coupling strength $\gamma=18.2m\omega_\parallel^2$. (c) Logarithmic negativity as a function of time for a chain with $N=800$ (solid line), $1000$ (dashed) and $1200$ (dotted) ions, for $d=17a$, $T=0$, $s=0.5$, and $\gamma=0.036\,m\omega_\textmd{ref}^2$. For each curve, the time is in units of $\omega_\parallel(N)^{-1}$, with $\omega_\parallel(N)=\omega_\textmd{ref}\sqrt{\log N}/N$ and $\omega_\textmd{ref}=2\pi\times 659.6$ kHz. Insets: corresponding spectral densities (in units of $m\omega_\parallel^2$) as a function of the frequency. The solid (dashed) line correspond to $J_+(\omega)$ ($J_-(\omega)$). The arrow indicates the defects frequency for the solid curves in (b) and (c), respectively \cite{laser_shift}.} 
\end{center}
\end{figure*}

We first analyze the spectrum of Hamiltonian $H_0$. For simplicity, we assume equally-spaced axial equilibrium positions with interparticle distance $a=z_{j+1}^{(0)}-z_j^{(0)}$, which can be found in the central region of long ion chains \cite{Morigi:2004} and in anharmonic potentials \cite{Duan,Champenois:2010}. Figure \ref{fig:spectrum} displays the spectra of the axial and transverse modes for $M\approx 2.87m$, which corresponds to In$^+$ ions embedded in a Ca$^+$ chain~\cite{Hayasaka}. Two degenerate normal mode frequencies for each transverse spectrum are separated from the respective transverse branch by a gap: If this gap is sufficiently large, these modes approximately coincide with the defects transverse vibrations. In this limit they are strongly coupled with each other via the Coulomb interaction, and weakly coupled to the rest of the chain. Their dynamics are governed by a beam-splitter type of interaction: The two defects can become entangled by the unitary evolution after preparing each transverse mode in a squeezed state \cite{BeamSplitter}. We verify this behaviour assuming that initially the chain is in a thermal state at temperature $T$ while the transverse modes are prepared in identical squeezed vacuum states along the $x$ direction with variances $\Delta x_{j_1,j_2}^2=x_0^2 \;e^{2s}/2$ and $\Delta p_{j_1,j_2}^2=p_0^2 \;e^{-2s}/2$. Here $x_0=\sqrt{\hbar/(M \Omega_{\perp})}$ is the size of the ground state of the defect oscillator with frequency $\Omega_{\perp}=\sqrt{U_\perp(M)/M}$, $p_0=\hbar/x_0$, and $s$ is the real-valued squeezing parameter \cite{BeamSplitter}. The defects' state remains Gaussian under the evolution given by the quadratic Hamiltonian, Eq.(\ref{eq:hamiltonian}), and is hence described by the first moments and the covariance matrix \cite{Adesso2007}. Entanglement between the defect modes is quantified by means of the logarithmic negativity, $E_N$, obtained from the symplectic eigenvalues of the partially transposed covariance matrix \cite{Vidal2002}.  The grey curves in Fig. \ref{fig:EN_dif_s} display $E_N$ as a function of the time $t$ for a fixed distance $d=7a$: $E_N$  grows with $t$ and increases with $s$. The defects become entangled also when they are initially not squeezed ($s=0$), which is due to the fact that the normal modes of the isolated frequencies include some small displacements of the other ions of the chain. Numerical simulations show that $E_N$ at a given time $t$ decreases with the strength of the direct Coulomb coupling, which scales with the distance as $1/d^3$.

We now demonstrate that entanglement between the defects can be generated and substantially enhanced by coupling with the axial phonons of the chain. For this purpose we consider an interaction which couples the defects' transverse and axial displacements. The dynamics are now given by Hamiltonian $H=H_0+H_I(t)$, where
\begin{equation}
  H_I(t)=\frac{\gamma(t)}{2}\left[(x_{j_1}-q_{j_1})^2+(x_{j_2}-q_{j_2})^2\right],
\end{equation}
with $\gamma(t)=\gamma\,\Theta(t)$ being an effective coupling strength and $\Theta(t)$ the Heaviside function. This coupling could be realized, for instance, using a standing-wave laser in the $x-z$ plane illuminating the defects in the Lamb-Dicke regime and with nodes at their equilibrium positions \cite{Cirac:1992}. For this geometry the motion along $y$ is decoupled and will be ignored from now on. At $t>0$ a displacement of each defect along $x$ excites a wave packet of axial phononic modes. 

We first note that these dynamics lead to thermalization of a single defect \cite{Rubin:1963,Kajari:2012}, but are significantly modified when both defects are coupled. In order to understand why, let us consider that the defects are symmetrically placed with respect to the trap center. By performing a change of coordinates, $q_j^{\pm}=(q_{N-j+1}\pm q_j)/\sqrt{2}$ and $X_\pm=(x_{j_1}\pm x_{j_2})/\sqrt{2}$, it becomes evident that the defects' transverse center-of-mass (COM) displacement, $X_+$, couples to the axial displacements $q_j^+$, and the same holds for $X_-$ and $q_j^-$. The action of the axial vibrations on the dynamics of the defects' collective variable can be characterized in terms of the spectral densities~\cite{Wolf:2011} 
$$J^{\pm}(\omega)=\pi\sum_{k=1}^{N/2}\frac{(\gamma_k^{\pm })^2}{2m\omega_k^{\pm}}\delta(\omega-\omega_k^{\pm})\,,$$ 
where $\gamma_k^\pm$ is the coupling strength of the coordinate $X_\pm$ to the $k$-th normal mode at eigenfrequency $\omega_k^\pm$  of the chain $q_j^{\pm}$. The inset of Fig. \ref{fig:EN_dif_s} displays the spectral densities $J_\pm(\omega)$ for $d=7a$. The appearance of frequency values $\omega_\ell^+(\omega_\ell^-)$, at which the spectral density $J_+$ ($J_-$) vanishes, is a signature of ``dark'' normal modes, namely, localized excitations which involve the collective variable $X^+$ ($X^-$) and which are decoupled from the rest of the chain. These zeroes equal the number of ions between the two defects, and each frequency $\omega_\ell^+$ ($\omega_\ell^-$) corresponds to a decoherence-free subspace for $X_+$ ($X_-$) when the defect frequency fulfills $\Omega_\perp\simeq \omega_\ell^+$ ($\Omega_\perp\simeq \omega_\ell^-$)~\cite{laser_shift}. Entanglement between the defects can thus be generated for an initial state in which both impurities are in a squeezed state while the rest of the chain is in a thermal state \cite{Note}. In fact, taking for instance $\Omega_\perp=\omega_\ell^-$, the dynamics leads to thermalization of the defects' COM motion, $X_+$, and destroys all initial correlations between the defects' COM and relative motion, while the relative coordinate preserves part of the initial squeezing. Sufficiently large squeezing $s$ and low temperatures $T$ lead to two-mode squeezing of the defects' transverse motion \cite{Paz:2008,Kajari:2012}, and hence to entanglement \cite{Reid2009}. 

The logarithmic negativity is determined after numerically evaluating the defects' dynamics, starting from the formal solution of the coupled Heisenberg equations of transverse, axial, and defect modes of the Hamiltonian $H$ \cite{Kajari:2012}. The solid curves in Fig. \ref{fig:EN_dif_s} display $E_N$ in a chain of $50$ ions, and for times over which finite-size effects are negligible. Entanglement builds up after a transient time and reaches values which are an order of magnitude larger than the values found in the absence of the coupling laser. It increases with the strength of the initial squeezing $s$, and we checked that it also increases when decreasing the temperature of the chain, which is consistent with the results of previous works \cite{Paz:2008,Wolf:2011,Kajari:2012}. We note that, while cooling a large chain to ultralow temperatures is a challenging task, the basic requirement for observing the dynamics predicted here is that the axial mode of the chain resonant with the defect frequency is prepared in the ground state. In a chain of trapped ions this could be realized, for example, by cooling the motion of the defect ions while the laser coupling transverse and axial displacement is switched on \cite{Sympathetic}. Noise will not significantly affect the predicted dynamics, provided that the dark mode is cold and protected from external noise sources \cite{Footnote:noise}. 

We now demonstrate that the created entanglement is not a finite-size effect, and is indeed independent of the chain size. Fig. \ref{fig:thermo_limit} displays the logarithmic negativity as a function of time for various chain sizes and a mutual distance of $d=17a$ between the defects. Here the COM motion is decoupled by tuning its frequency to a value at which the corresponding spectral density vanishes (see inset). The curves are displayed for times that are shorter than the revival time and have been rescaled by the size-dependent frequency $\omega_\parallel=\omega_\textmd{ref}\sqrt{\log N}/N$, where $\omega_\textmd{ref}$ is a constant and the size-dependent factor warrants that the interparticle distance at the chain center is independent of $N$ (therefore keeping constant the cutoff frequency) \cite{Morigi:2004}. One observes that the logarithmic negativity oscillates about a (quasi) stationary state, whose value is independent of the chain size. For increasing chain sizes the revival time, and thus the time window over which this entanglement is found, correspondingly increases.

\begin{figure}[t]
\begin{center}
\includegraphics[width=0.93\columnwidth]{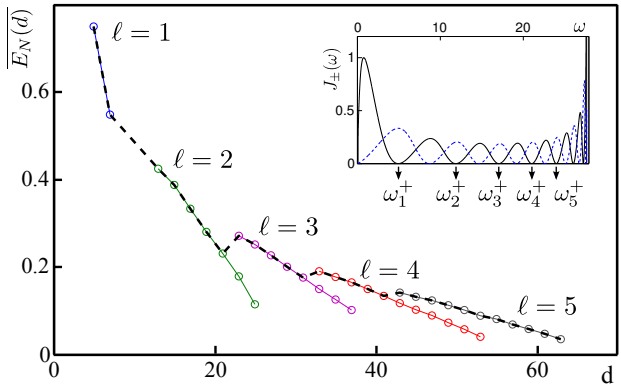}
\caption{(Color online) Steady-state entanglement as a function of the defects' mutual distance. Solid lines connect $\overline{E_N}$ for COM decoupling via the $\ell$-th zero of $J^+(\omega)$. Dashed line shows the maximun $\overline{E_N}$. Parameters are set to $N=800$, $T=0$, $s=0.5$ and coupling $\gamma=3230m\omega_\parallel^2$. Inset: Spectral densities for $d=15a$. Arrows indicate corresponding frequencies $\omega^+_\ell$.\label{fig:scaling}}
\end{center}
\end{figure}

Let us finally show that the entanglement generated by the interaction with the chain slowly decays with the mutual distance. Figure \ref{fig:scaling} displays the mean value of the logarithmic negativity, $\overline{E_N}$, averaged over the time in which it regularly oscillates, as a function of the mutual distance when the COM is decoupled. The different points correspond to different zeroes of the spectral density: optimal entanglement is thus achieved by suitably tuning the frequency of the impurity, so that it matches the frequency of an optimal dark normal mode. Steady-state entanglement is found on time scales $t_0$ of the order of $d/c_s$, where $c_s\simeq \omega_\| a$ is the sound velocity \cite{Morigi:2004}. For the parameters and the chains here considered $t_0\sim100\mu s$. At large $d$, $E_N$ decays linearly with the distance, which is confirmed by a systematic analysis performed on a model with nearest-neighbor interactions \cite{Kajari:2013}. 

The resulting entanglement can be measured by extending the method developed in Ref.~\cite{Tufarelli2011} to two Gaussian modes. The procedure consists of coupling the transverse oscillation of each defect with an ancillary qubit, such as an electronic transition of the defect ion. This can be done using the Hamiltonian $H^{int}_{j=j_1,j_2}=\hbar g_j(t)\sigma_j^z(a_j e^{-i\Omega_j\bar t}+a^\dagger_j e^{i\Omega_j\bar t})$, where $a_j=(x_j/x_0+{\rm i}p_{xj}/p_0)/\sqrt{2}$ annihilates a phonon of the defect oscillator at frequency $\Omega_{\perp}$, $g_j$ denotes the Rabi coupling with the internal transition, and $\sigma_j^{x,y,z}$ are the Pauli operators. The expectation value $\left<T\right>=\left<\sigma^x\otimes\sigma^x-\sigma^y\otimes\sigma^y+i\sigma^x\otimes\sigma^y+i\sigma^y\otimes\sigma^x\right>$ gives the characteristic function $\chi(\beta_1,\beta_2)$ of a two-mode system, where $\beta_{1,2}=2i\int_0^{t}dt' g_{j_1,j_2}(t')e^{i\Omega_{j_1,j_2}t'}$. The reconstruction of the oscillators state and thus of its entanglement properties is granted by properly designed coupling profiles. Note that it is sufficient to probe the characteristic function of Gaussian states close to the phase-space origin since they are characterised by their first and second moments.

Our study sheds light on the role of the reservoir in establishing quantum correlations. It further identifies the basic requirements for entangling two distant objects which interact with a bulk exhibiting long-range order. Here, non-local, decoherence-free subspaces appear which are associated with normal modes involving the defects vibrations. In principle, it is thus possible to entangle arbitrarily distant objects via a common bath. An intrinsic limitation to such dynamics is the fact that the required time linearly scales up with the distance, while the zeroes of the dark normal modes become denser and denser. A further challenge is the realization of arbitrary large regular lattices. Tuning the defects' frequency is the key for entanglement generation in other physical platforms. These dynamics can also be observed, for instance, in dipolar quantum gases in the Mott insulator phase in an optical lattice, where the optical traps can be engineered in order to tune the frequency of the transverse oscillations of different species \cite{OpticalLattices,Lahaye:2009}. Other examples are cavity arrays in circuit QED \cite{CircuitQED}  and optomechanical systems as in \cite{Marquardt:2012}, where the frequencies can be controlled by means of proper designs and the use of refractive media.

The authors ackowledge discussions with C. Cormick, J. Brito, G. De Chiara, C. Kurz, and E. Lutz, and support by the European Commission (IP AQUTE, STREP PICC), the German Research Foundation and the Irish Research Council through the Embark Initiative RS/2009/1082.


\begin{thebibliography}{99}
\bibitem{Note}
Although this argument is based on the reflection symmetry about the chain center, it is valid for any positioning of the defects as long as they are far away from the edges is large and their mutual distance is much smaller than the chain size.
\end{thebibliography}
\end{document}